\documentclass[useAMS,usenatbib]{mn2e}
\usepackage{graphicx}

\title{The solar system test for the general modified gravity theories}
\author[ Chan \& Lee]{ Man Ho Chan \thanks{chanmh@eduhk.hk} and Chak Man Lee
\\ Department of Science and Environmental Studies, The Education University of Hong Kong, Tai Po, Hong Kong}

\begin{document}

\date{Accepted XXXX, Received XXXX}

\pagerange{\pageref{firstpage}--\pageref{lastpage}} \pubyear{XXXX}

\maketitle

\label{firstpage}

\date{\today}

\begin{abstract}
In the past few decades, various versions of modified gravity theories were proposed to mimic the effect of dark matter. Compared with the conventional Newtonian or relativistic dynamics, these theories contain some extra apparent force terms in the dynamical equations to replace the role of dark matter. Generally speaking, the extra apparent force terms usually scale with radius so that the effect would be significant only on large scale to explain the missing mass in galaxies or galaxy clusters. Nevertheless, the apparent effect may still be observable in small structures like the solar system. In this article, we derive analytic general formulae to represent the contribution of the precession angle of the planets in the solar system due to the general modified gravity theories, in which the extra apparent force terms can be written in a power law of radius $r$ or an exponential function in $r$. We have tested three popular modified gravity theories, the Modified Newtonian Dynamics (MOND), the Emergent Gravity (EG), and the Modified Gravity (MOG). In particular, based on the solar system data, we have constrained the parameters involved for two popular general interpolating functions used in MOND. Our results can be generally applied to both of the modified inertia and modified gravity versions of MOND.
\end{abstract}


\section{Introduction}
Dark matter has been postulated to account for the missing mass observed in our universe. However, no positive signal of dark matter has been detected so far in the collider experiments \citep{Abecrcrombie}, direct-detection experiments \citep{Amole,Aprile}, and indirect-detection experiments \citep{Ackermann,Chan,Chan2,Chan3}. Therefore, some theories suggest that current Newtonian or relativistic dynamics has to be corrected in large-scale structures so that the extra correction terms in the dynamical equations can mimic the effect of dark matter \citep{Bertone}. These theories are commonly known as modified gravity theories.

In particular, one of the earliest versions is the Modified Newtonian Dynamics (MOND) proposed by Milgrom \citep{Milgrom}. It suggests a significant change in the Newton's second law when the acceleration is smaller than a critical value $a_0 \sim 10^{-8}$ cm/s$^2$. Later, Moffat proposes another theory called the Modified Gravity(MOG), which suggests a change in the Newtonian gravitational potential when the structural radius $r$ is larger than kpc \citep{Moffat}. Recently, Verlinde has formulated a new theory called the Emergent Gravity (EG) \citep{Verlinde}. The EG can directly connect the apparent missing mass with the real baryonic mass and it predicts the existence of a universal acceleration scale $a_0$ which is consistent with the proposal of MOND. As highlighted and discussed in \citet{Bertone}, they are three of the most popular modified gravity models which can satisfactorily explain the missing mass (i.e. dark matter) observed in galaxies \citep{Sanders,Milgrom2,Green,Davari,Brouwer}.

Generally speaking, the apparent extra gravitational effects of these modified gravity theories are very small when the scale of the structure is smaller than a galaxy. However, these effects are not completely negligible in small structures. It has been shown that some effects of modified gravity could be measurable in the solar system. For example, some previous studies have investigated the observable effects due to an additional acceleration term, such as the effect on spacecraft (the Pioneer anomaly) \citep{Milgrom3} and the effects on planetary motions \citep{Sanders3,Milgrom4}. In particular, the orbital precession of planets in the solar system is one of the most important indicators to examine modified gravity theories. Some previous studies have discussed the possible effects \citep{Sanders3,Iorio,Blome,Blanchet,Famaey2,Hees2,Hees}. Nevertheless, most of the previous discussions usually focus on particular forms of the modified gravity. In this article, we revisit the solar system test and provide a more comprehensive discussion on the results. We derive general analytic formulae of orbital precession angle for the general modified gravity theories, in which the extra apparent force term is a power law in radius $r$ or an exponential function in $r$. Based on the analytic formulae derived, we have tested three popular versions of modified gravity theories. In particular, we have obtained constraints for the parameters involved for two very popular general interpolating functional forms used in MOND. Our results can be generally applied to both of the modified inertia (for low eccentricity orbits) and modified gravity versions of MOND. 

\section{The general precession formula}
In the Schwarzschild model, the spherical symmetric space-time metric near the Sun can be well-described as
\begin{equation}
ds^2=A(r)c^2dt^2-B(r)dr^2-r^2(d\theta^2+\sin^2\theta d\phi^2),
\label{metric}
\end{equation}
where $(r, \theta,\phi)$ are the spherical coordinates, $A(r)=1-r_s/r$ with $r_s=2GM_{\odot}/c^2$, and $B(r)=1/A(r)$. The motion of a planet, on the fixed plane $\theta=\pi/2$ without loss of generality, is then given by
\begin{equation}
\frac{d^2u}{d\phi^2}+u=\frac{GM_{\odot}}{L^2}+3\frac{GM_{\odot}}{c^2}u^2,
\label{trajectory}
\end{equation}
where $u=1/r$ and $L$ is the angular momentum. The second term on the right-hand side represents the general relativistic effect due to the Sun.

In the absence of the general relativistic term, Eq. (\ref{trajectory}) has the analytic solution
\begin{equation}
u=u_0(1+e\cos\phi),
\end{equation}
where $u_0=GM_{\odot}/L^2=[(a(1-e^2))]^{-1}$, $a$ and $e$ are the semi-major axis and the eccentricity respectively. Nevertheless, when the general relativistic term is included, the orbit would undergo prograde precession.

Suppose that a modified gravity theory contains an extra apparent force term in the equations of motion, in which the force term is in the power-law functional form of $r$. This extra apparent force term can be viewed as the gravitational effect due to an extra apparent dynamical mass with a power law in $r$:
\begin{equation}
M_{\rm ex}(r)=Kr^n,
\label{powerform}
\end{equation}
where $K$ and $n$ are constant parameters. Therefore, neglecting the mass of planets and dwarf planets, the total apparent enclosed dynamical mass experienced by a planet inside the radius $r$ of the solar system would be $M_{\rm dyn}(r)=M_{\odot}+M_{\rm ex}(r)$. Generally speaking, the extra apparent gravitational force term is very small compared with the solar mass (i.e. $M_{\rm ex}(r) \ll M_{\odot}$) in the solar system. Therefore, this extra term is a perturbation term. To include this extra apparent force due to the modified gravity theory, we can replace the term $GM_{\odot}/L^2$ in Eq.(\ref{trajectory}) by $GM_{\rm dyn}(r)/L^2$. To keep the first order of perturbation, we can neglect the extremely small term $3GM_{\rm ex}(u)u^2/c^2$. The equation of motion can be written explicitly as
\begin{equation}
\frac{d^2u}{d\phi^2}+u=\frac{GM_{\odot}}{L^2}+\frac{GK}{L^2}u^{-n}+3\frac{GM_{\odot}}{c^2}u^2.
\label{trajectory2}
\end{equation}

Let $u=u_0+\Delta u$ with $\Delta u\ll u_0$, where $u_0$ is a constant and $\Delta u$ is a function of $\phi$. By expanding all terms of Eq. (\ref{trajectory2}) in Taylor series up to the first order of $(\Delta u/u_0)$, we get
\begin{eqnarray}
\frac{d^2}{d\phi^2}\Delta u&\approx&\left[\frac{GM_{\odot}}{L^2}+\frac{G}{L^2}\frac{K}{u_0^n}+\frac{3GM_{\odot}}{c^2}u_0^2-u_0\right]\nonumber\\
&&-\left[1-\frac{6GM_{\odot}u_0}{c^2}+
\frac{G}{L^2}\frac{Kn}{u_0^{n+1}}
\right]\Delta u.
\label{trajectory3}
\end{eqnarray}
We choose the arbitrary $u_0$ such that the first term on the right-hand side is zero. Then Eq. (\ref{trajectory3}) becomes
\begin{equation}
\frac{d^2}{d\phi^2}\Delta u=-(1-\alpha)^2\Delta u,
\end{equation}
where
\begin{equation}
(1-\alpha)^2\approx (1-2\alpha) \approx \left[1-\frac{6GM_{\odot}u_0}{c^2}+\frac{G}{L^2}\frac{Kn}{u_0^{n+1}} \right].
\end{equation}
The general solution of Eq.~(7) $\Delta u=\Delta u(\phi=0)\cos(1-\alpha)\phi$ can give the general form of an ellipse with precession as
\begin{equation}
r=\frac{r_0}{1+e\cos(1-\alpha)\phi}.
\end{equation}
Here, since $M_{\rm ex}(r) \ll M_{\odot}$ in the solar system, we have $u_0=1/r_0 \approx GM_{\odot}/L^2 \approx [a(1-e^2)]^{-1}$. Therefore, the analytic form of the precession angle per one period is
\begin{equation}
\Delta \phi=2 \pi \alpha=\frac{6\pi GM_{\odot}}{a(1-e^2)c^2}- \frac{\pi Kn}{M_{\odot}}[a(1-e^2)]^n.
\end{equation}
Note that the general relativistic precession term $6\pi GM_{\odot}/a(1-e^2)c^2$ is prograde while the precession term $\pi Kn[a(1-e^2)]^n/M_{\odot}$ due to modified gravity is retrograde. Moreover, not only for modified gravity theories, the analytic precession angle formula can also be applied for examining the effect on precession due to dark matter or extended mass distribution in which the enclosed mass function is a power law in $r$.

\section{Testing the modified gravity theories}
In the followings, we will test three popular modified gravity theories: MOND, EG and MOG. These theories contain an apparent dynamical mass term in the equation of motion in a power-law form of $r$ to account for the missing mass observed.

\subsection{Modified Newtonian Dynamics (MOND)}
The apparent gravitational acceleration $g$ in MOND can be given by
\begin{equation}
g=\nu(g_b/a_0)g_b,
\end{equation}
where $\nu(x)$ is the MOND interpolating function (IF), $a_0 \approx 1.2 \times 10^{-8}$ cm/s$^2$ \citep{McGaugh} and $g_b=GM_{\odot}/r^2$. There is no theoretical predicted form for $\nu(x)$, but a simple IF is widely used in recent literature \citep{Famaey,Sanders2,McGaugh3,Chae,Wang}:
\begin{equation}
\nu(x)=\frac{1}{2}+\sqrt{\frac{1}{4}+\frac{1}{x}}.
\end{equation}
As discussed in \citet{Famaey}, this IF provides a less sudden transition from the Newtonian to the MONDian regime than other standard functions, which can give a better agreement with the galactic rotation curve data, and a more realistic mass-to-light ratio in galaxies \citep{Sanders2}. Recent studies have shown that this simple IF can also give good agreements with observational results in the Milky Way galaxy \citep{McGaugh3} and elliptical galaxies \citep{Chae,Sanders2,Chae2}.  Nevertheless, in order to give a more comprehensive analysis, we consider a more general functional form of the IF (hereafter called general IF) \citep{Famaey2}:
\begin{equation}
\nu(x)=\left[\frac{1+(1+4x^{-p})^{1/2}}2\right]^{1/p},
\end{equation}
where $p>0$ is a free parameter. When $p=1$, the above general IF would reduce to the simple IF.

Within the solar system, we have $g_b \gg a_0$. To keep the first order of $a_0/g_b$, we can obtain $\nu(g_b/a_0) \approx 1+p^{-1}(g_b/a_0)^{-p}$. Therefore, the apparent extra mass due to MOND is
\begin{equation}
M_{\rm ex}(r) \approx \frac{1}{p} \left(\frac{GM_{\odot}}{a_0} \right)^{-p}M_{\odot}r^{2p}.
\end{equation}
Hence, we get $K=p^{-1}(GM_{\odot}/a_0)^{-p}M_{\odot}$ and $n=2p$ for the general IF.

Moreover, we also consider another popular IF in MOND (hereafter called $\delta$-family IF) \citep{Famaey2,Dutton}:
\begin{eqnarray}
\nu(x)&=&[1-\exp(-x^{\delta/2})]^{-1/{\delta}}\nonumber\\
&\approx&1+\frac{1}{\delta}\exp(-x^{\delta/2})\,\,\,\,\,\,{\rm for}\,\,x\gg 1,
\end{eqnarray}
where $\delta>0$ is a free parameter. This functional form of IF has been widely used recently because it can give excellent agreements with rotation curve data (with $\delta=1$) for many galaxies \citep{McGaugh4,Lelli,Li}. The corresponding extra mass term is given by
\begin{eqnarray}
M_{\rm ex}&=&\frac{1}{\delta}\exp(-x^{\delta/2})M_{\odot}\nonumber\\
&=&\frac{1}{\delta}\sum_{k=0}^{\infty}
\frac{1}{k!}\left[-\left(\frac{GM_{\odot}}{a_0}\right)^{\delta/2}\frac{1}{r^{\delta}}\right]^k M_{\odot}.
\end{eqnarray}
By using the same technique in deriving Eq.~(6), we have
\begin{eqnarray}
\frac{d^2}{d\phi^2}\Delta u&\approx&
\left[\left(1+\frac{1}{\delta}\right)\frac{GM_{\odot}}{L^2}+\frac{3GM_{\odot}}{c^2}u_0^2-u_0\right.\nonumber\\
&&\left.
+\frac{GM_{\odot}}{L^2}
\sum_{k=1}^\infty \frac{1}{k!}\frac{1}{\delta}\left(-\frac{GM_{\odot}}{a_0}\right)^{k\delta/2}u_0^{k\delta}
\right]\nonumber\\
&&-\left[1-\frac{6GM_{\odot}u_0}{c^2}\right.\nonumber\\
&&\left.-\frac{GM_{\odot}}{L^2}
\sum_{k=1}^\infty \frac{1}{k!}\left(-\frac{GM_{\odot}}{a_0}\right)^{k\delta/2}ku_0^{k\delta-1}
\right]\Delta u.\nonumber\\
\label{trajectory1}
\end{eqnarray}
Therefore, the precession angle contributed by the extra force term for the $\delta$-family IF is then given by
\begin{eqnarray}
\Delta\phi&=& \frac{6\pi GM_{\odot}}{a(1-e^2)c^2}\nonumber\\
&&+
\sum_{k=1}^\infty \frac{\pi}{(k-1)!}\left(-\frac{GM_{\odot}}{a_0}\right)^{k\delta/2}\left[\frac{1}{a(1-e^2)}\right]^{k\delta}\nonumber\\
&=&\frac{6\pi GM_{\odot}}{a(1-e^2)c^2}-{\pi} f\exp(-f),
\end{eqnarray}
where
\begin{equation}
f=\left(\frac{GM_{\odot}}{a_0}\right)^{\delta/2}\frac{1}{a^{\delta}(1-e^2)^{\delta}}.
\end{equation} 
This is the first time to obtain the analytic general precession formulae for the general IF and the $\delta$-family IF.

By using the analytic general precession formulae, we can calculate the predicted precession angles for 6 solar system planets (the data for Uranus and Neptune are not reliable). Here, the precession angle we compared refers to the residual precession angle after considering the known Newtonian effects, such as the gravitational tugs of other solar bodies, quadruple moment of the Sun and the Lense-Thirring precession effect \citep{Park}. Therefore, the residual precession angle includes the General Relativity effect and the extra effect due to modified gravity. To constrain the IF, we compare the predicted precession angles with the observed residual precession angles (after eliminating the known Newtonian effects) for 6 solar system planets. For Mercury, latest results based on the Cassini experiment \citep{Bertotti} and the data from Messenger spacecraft \citep{Fienga} have provided a very stringent constraint on the allowed extra precession angle (in arcsec per century): $-5.87537\times 10^{-4}<\Delta \phi-42.98<2.96635 \times 10^{-3}$ \citep{March}, where $42.98$ arcsec per century is the general relativistic precession angle. Besides, we also compare with the data of the observed precession angles shown in \citet{Nyambuya} for the other 5 planets. The basic information and the observed precession angles for the 6 planets considered are shown in Table 1 and Table 2 respectively. 

We first perform the analysis for the simple IF based on our general analytic precession formula (see Table 3). We can see that almost all predictions (except Venus) do not agree with the observed limits (i.e. observed value $-$ general relativistic term). Therefore, the MOND with the simple IF (the simple version of MOND) does not pass the solar system test. For the general IF and the $\delta$-family IF, we can constrain the parameters $p$ and $\delta$ involved in the functional forms. We mainly consider the data of Mercury as it can give the most stringent constraints. By using Eq.~(10) and Eq.~(18), the allowed values of $p$ and $\delta$ are $p \ge 1.22$ and $\delta \ge 0.33$ respectively. In other words, $p<1.22$ and $\delta<0.33$ are ruled out by the solar system test. In view of these results, the so-called standard IF (i.e. the general IF with $p=2$) used in some previous studies \citep{vandenBosch,Dutton} can pass the solar system test, although some studies have shown that it gives poorer fits with rotation curves for some galaxies \citep{Famaey}. Moreover, the $\delta$-family of IF with $\delta=1$ used in many recent studies \citep{McGaugh4,Lelli,Li,Dutton} can also pass the solar system test. 

Note that there are two different versions of MOND in general: modified inertia version and modified gravity version. For the modified gravity version, \citet{Blanchet,Hees2} have already shown that only very large values of $p$ and $\delta$ are allowed, based on the data of Saturn precession and some galactic rotation curves. However, these constraints cannot apply to modified inertia version of MOND \citep{Hees2}. The derivations of our general analytic precession formulae are based on the general relation in MOND so that the formulae can also be applied to constrain the interpolating functions involved for the modified inertia version of MOND. Therefore, our results can provide critical tests to the modified inertia version of MOND \citep{Milgrom5} based on the solar system data.  

\subsection{The Emergent Gravity (EG)}
In the EG model, baryonic matter displaces dark energy to give an additional elastic force \citep{Verlinde,Verlinde2}. The apparent extra mass term in the solar system is given by
\begin{eqnarray}
M_{\rm ex}(r)&=&\sqrt{\frac{a_Mr^2}{6G}\frac{d}{dr}[rM_{\odot}(r)]}\nonumber\\
&=&\sqrt{\frac{a_MM_{\odot}}{6G}}r,
\end{eqnarray}
where $a_M\approx cH_0\approx 7\times 10^{-8}~{\rm cm/s^2}$. Therefore, we have $K=\sqrt{a_MM_{\odot}/6G}$ and $n=1$. We regard this as the EG1 model. Again, by using the analytic precession formula Eq.~(10), we can calculate the predicted precession angles for the 6 planets (see Table 3). We can see that the predicted values are much larger than the observed limits for all planets. Since there is no free parameter in the formalism, the EG1 model definitely fails the solar system test. This result is consistent with the conclusion in \citet{Hees}, although the degree of discrepancy between the predicted values and observed limits is different.

On the other hand, a recent study suggests that the apparent gravitational acceleration in the EG theory should be given by another relation $g=\sqrt{g_b^2+g_D^2}$ (the EG2 model), where $g_D^2=a_Mg_b/6$ \citep{Yoon}. This relation can be translated to $g \approx g_b+a_M/12$, which is similar to the simple version of MOND. The only difference is that the acceleration constant becomes $a_M/12 \approx 5.8 \times 10^{-9}$ cm/s$^2$. By comparing the calculated precession values and the observed limits, we can see that this version has shown significant discrepancies in the predicted precession angles for the Earth, Mars, Jupiter and Saturn (compare Table 2 and Table 3). Therefore, it can't pass the solar system test either.

\subsection{The Modified Gravity (MOG)}
The MOG is a covariant theory of gravity which assumes a massive vector field coupled to baryonic matter. The Yukawa-type modification of the Newton's law of gravitation can be written as \citep{Moffat}
\begin{equation}
g=\frac{GM_{\odot}}{r^2}\{1+\alpha[1-(1+\mu r)\exp(-\mu r)] \},
\end{equation}
where $\alpha=8.99 \pm 0.02$ and $\mu=0.054 \pm 0.005$ kpc$^{-1}$ are the constant parameters constrained by the Milky Way data \citep{Davari}. In the solar system, since $\mu r \ll 1$, the apparent extra mass term can be written as
\begin{equation}
M_{\rm ex}(r) \approx \alpha \mu^2M_{\odot}r^2.
\end{equation}
Therefore, we have $K=\alpha \mu^2M_{\odot}$ and $n=2$. By using the analytic precession formula Eq.~(10), we find that the predicted precession angles are of the order of $10^{-11}-10^{-10}$ arcsec per century (see Table 3), which are too small to violate the observed limits. In other words, the MOG can safely pass the solar system test.

\section{Discussion}
In this article, we have first derived the analytic general precession angle formulae for the general modified gravity theories, in which the extra force terms can be written in a power law in $r$ or an exponential function in $r$. This is the first time to obtain these general analytic formulae for general modified gravity theories, especially for the general IF and the $\delta$-family IF in MOND theory. The general analytic formulae obtained in this study can provide an easy and systematic test for different general IF in MOND and different modified gravity theories. On the other hand, our general formulae can also be used for examining or estimating the effect on the orbital precession angle by any dark matter distribution or extended mass distribution with a power law in $r$ (e.g. the isothermal density profile) at the Galactic Centre \citep{Chan7}.

By applying the derived formulae, we have specifically tested three popular modified gravity theories highlighted in \citet{Bertone}: MOND, EG and MOG. We have shown that the simple version of MOND (the simple IF) and the EG cannot pass the solar system test while the MOG can safely pass the test. In particular, for the MOND theory, we have shown that $p \ge 1.22$ and $\delta \ge 0.33$ are allowed for the two popular general IF used in MOND. In fact, previous studies have already shown that only large values of $p$ and $\delta$ are allowed for the modified gravity version of MOND \citep{Blanchet,Hees2}. Here, since our formulae can also be applied to the modified inertia version of MOND, our limits on $p$ and $\delta$ can provide important constraints for the parameters involved in the modified inertia version of MOND. 

For the EG theory, we have examined two possible variations in the formalism (the EG1 and EG2 models). However, both of them have predicted very large precession angles for planets which violate the observed limits. Since there is no free parameter given in the EG theory, it should be definitely falsified, unless adding some other possible auxiliary hypothesis (e.g. screening mechanism) to the theory. For the MOG theory, the apparent extra gravity on the scale of the solar system is very small. Therefore, the observable effect is nearly negligible.

Generally speaking, a good modified gravity theory should be able to explain the missing mass or dark matter observed from small to large scales. Therefore, the tests from the solar system (small scale), galaxies (medium scale) and galaxy clusters (large scale) are all necessary to pass. Although the MOND theory can somewhat pass the solar system test (with the standard IF or the $\delta$-family IF with $\delta \ge 0.33$) and give generally good agreements with the data in galaxies \citep{McGaugh}, it cannot give satisfactory results in galaxy clusters \citep{Sanders,Angus}. Also, whether there exists a universal acceleration scale suggested in MOND is quite controversial \citep{Wang,McGaugh2,Rodrigues,Chan4,Chan5,Chan6}.  Moreover, the relativistic version of MOND (the TeVeS theory) is also greatly challenged by recent gravitational wave detection \citep{Gong}, although some studies have demonstrated a new paradigm of TeVeS-like theories to satisfy the gravitational wave constraints \citep{Skordis}. Therefore, it is still controversial to regard MOND as a good modified gravity theory. 

For the EG theory, although it has gained some credence in galaxies and galaxy clusters \citep{Brouwer,Tortora,Tamosiunas}, it gives a very large discrepancy in the solar system test. Therefore, the EG model in current formalism should be ruled out. For the MOG model, so far it can give good agreements in the solar system (our work) and galaxies \citep{Green,Davari,Moffat2}. However, for galaxy clusters, there are some positive \citep{Moffat3} and negative evidence \citep{Martino} on the MOG model. Therefore, more observational data and analyses are required to give a better assessment or conclusion on the MOG model.

\begin{table}
\caption{Semi-major axis $a$, eccentricity $e$ and period of the six solar planets \citep{Nyambuya}.}

\begin{tabular}{ |l|c|c|c|}
 \hline\hline
 Planet       & Semi-major axis $a$ (AU)      &  Eccentricity $e$ & Period (days) \\

  \hline
 Mercury & 0.3871 & 0.206 & 88.97 \\
 Venus & 0.7233   & 0.007 & 224.70 \\
 Earth & 1.0000   &  0.017 & 365.26 \\
 Mars &  1.5237  &   0.093 & 686.98 \\
 Jupiter & 5.2034   & 0.048 & 4332.59 \\
 Saturn & 9.5371   &  0.056 & 10759.22 \\
  \hline\hline
\end{tabular}
\end{table}

\begin{table}
\caption{The precession angle (arcsec per century) contributed by the general relativistic term \citep{Park} and the observed residual precession angle (adopted from \citet{Nyambuya,March}) for the six solar planets.}

\begin{tabular}{ |l|c|c|}
 \hline\hline
 Planet                   & General relativistic term & Observed \\
  \hline
 Mercury & 42.9799 & $42.9799^{+0.0030}_{-0.0006}$ \\
 Venus& 8.628   & $8 \pm 5$ \\
 Earth& 3.841   & $5 \pm 1$ \\
 Mars&  1.351  &  $1.3624 \pm 0.0005$ \\
 Jupiter& 0.0623    & $0.070 \pm 0.004$ \\
 Saturn & 0.0137    & $0.014 \pm 0.002$ \\
  \hline\hline
\end{tabular}
\end{table}

\begin{table}
\caption{The precession angle (arcsec per century) contributed by the extra force terms from different modified gravity theories (simple IF in MOND, EG1, EG2 and MOG models) for the six solar planets. The negative signs indicate the retrograde precession of the orbits.}

\begin{tabular}{ |l|c|c|c|c|}
 \hline\hline
                    & MOND       &  EG1      & EG2        & MOG \\
                    & (simple IF) &          &            &     \\
  \hline
 Mercury&  -1.496 & -13988 & -0.727 &-$4.56\times10^{-11}$  \\
 Venus& -2.229 & -10684  & -1.084 &-$6.79\times10^{-11}$  \\
 Earth& -2.620 & -9085 & -1.273 &-$7.98\times10^{-11}$  \\
 Mars&  -3.180 & -7298 & -1.546 & -$9.69\times10^{-11}$ \\
 Jupiter& -5.956 & -3977 & -2.895 &-$1.81\times10^{-10}$ \\
 Saturn & -8.044  & -2933  & -3.910 &-$2.45\times10^{-10}$ \\
  \hline\hline
\end{tabular}
\end{table}

\section{Acknowledgements}
We thank the anonymous referee for useful constructive feedback and comments. The work described in this paper was partially supported by the Seed Funding Grant (RG 68/2020-2021R) and the Dean's Research Fund of the Faculty of Liberal Arts and Social Sciences, The Education University of Hong Kong, Hong Kong Special Administrative Region, China (Project No.: FLASS/DRF 04628).

\section{Data availability statement}
The data underlying this article will be shared on reasonable request to the corresponding author.

\label{lastpage}

\end{document}